# Sparsity Measure of a Network Graph: Gini Index

Swati Goswami[ab*], C. A. Murthy[a], Asit K. Das[b]

[a]Machine Intelligence Unit, Indian Statistical Institute, 203 B. T. Road, Kolkata - 700108, India

[b]Department of Computer Science and Technology, Indian Institute of Engineering Science and Technology, Shibpur, Howrah – 711103, India

## Abstract

This article examines the application of a popular measure of sparsity, Gini Index, on network graphs. A wide variety of network graphs happen to be sparse. But the index with which sparsity is commonly measured in network graphs is edge density, reflecting the proportion of the sum of the degrees of all nodes in the graph compared to the total possible degrees in the corresponding fully connected graph. Thus edge density is a simple ratio and carries limitations, primarily in terms of the amount of information it takes into account in its definition. In this paper, we have provided a formulation for defining sparsity of a network graph by generalizing the concept of Gini Index and call it sparsity index. A majority of the six properties (viz., Robin Hood, Scaling, Rising Tide, Cloning, Bill Gates and Babies) with which sparsity measures are commonly compared are seen to be satisfied by the proposed index. A comparison between edge density and the sparsity index has been drawn with appropriate examples to highlight the efficacy of the proposed index. It has also been shown theoretically that the two measures follow similar trend for a changing graph, i.e., as the edge density of a graph increases its sparsity index decreases. Additionally, the paper draws a relationship, analytically, between the sparsity index and the exponent term of a power law distribution, a distribution which is known to approximate the degree distribution of a wide variety of network graphs. Finally, the article highlights how the proposed index together with Gini index can reveal important properties of a network graph.

## 1. Introduction

When a group of people interact among themselves for some purpose through some channel, an interaction graph is born. The nodes of such a graph represent the people in the group, the edges represent the connections among individuals in the group and the edge weights represent the strength of such interactions. Of late, studying properties of interaction graphs has attracted attention of researchers from multiple disciplines like sociology, cognitive science, computer science, mathematics etc., each field having its own point of view to look at such graphs to solve problems related to its own specific field or problems which are multidisciplinary in nature. Interaction graphs can be studied from the records of communication among people in the form of voice calls, SMS messages, emails, social network interactions and so on. More generally, when any group of entities, animate or inanimate, are involved in some kind of relationships among themselves, a network graph can be used to reflect such relationships. Co-authorship relationship in an academic community, friend relationship in a university environment, protein-protein interactions in biological setting have all been studied in the literature using network graphs.

One of the distinguishing properties of a large network graph is its sparsity, which is an indication of the extent of its deviation from a fully connected graph. The more the deviation the higher is the sparsity.

The property of sparsity of a graph turns out to be a fundamental property looking at its various applications in constructing algorithms in different scenarios (graph clustering [3], graph based collaborative filtering [12] and so on). But there is no uniform definition of sparsity in the literature [13]. Not only that, the term sparsity carries different sense while being applied to different settings. In graph theoretic literature sparsity is a measure of the extent of a graph's deviation from the corresponding fully connected graph; but in many applied fields like signal processing or economics or sociology, sparsity is a measure indicating relative diversity among related entities with respect to a certain quantity of interest. If we do not get into the sparse family of graphs, which leads to a different direction outside of the scope of the present paper, and restrict ourselves to sparsity of a single graph, by saying that a graph is sparse we mean that the corresponding adjacency matrix is sparse, i.e., majority of the elements of the adjacency matrix are zeroes (or close to zeroes). But when we talk about sparsity in signal processing [21] [22], or, sparsity in wealth distribution in economics [14], we are more interested in it in the form of a measure indicating the relative distribution of a resource among the constituent entities. This subtle difference in its sense while being applied into different settings has led us to examine it afresh, in the context of network graphs, looking for a unifying measure.

---

[*] Corresponding Author. E-mail address: swati.goswami2000@gmail.com; Tel: +91 9836360340; Fax: +91 33 25783357. The author is on leave from her employment at the Research and Innovation unit of the Tata Consultancy Services, Kolkata, India

In the following sections of this document, some of the existing measures of sparsity of a network graph have been reviewed; a variant of Gini Index [10] has been proposed as a new measure of sparsity of a network graph with an assessment of how the new measure fares compared to the existing measures of sparsity with respect to certain properties desirable of such measures [13][21]. As more and more social phenomena are now being modeled for analysis using network graphs, it may be of interest to see how the property of sparsity relates with another important property of a network graph, viz., its degree distribution. As pointed out by Muchnik et al. [17] distribution of the number of ties of a person, as represented by the degree of a node in a network graph, has been shown to follow power law distribution for a large number of social networks and this scale-free nature is considered to be one of the vital properties of social networks. Considering the significance of degree distribution in characterizing and analysis of a network graph, its relationship with sparsity has been examined and the findings have been presented in section V. The article has been concluded with a discussion on applications of the suggested sparsity measure for network graphs and some direction in which further work on related topic can proceed.

## 2. *Review of existing measures of sparsity of a network graph*

To judge whether a network is dense or sparse, as a rule of thumb it is said that it is dense or sparse if the number of edges in it is quadratic or linear in the number of nodes [18]. But when the question of how to define the sparsity of a network graph is posed, a precise answer is often found missing in the literature. The property of sparsity is nevertheless a crucial one, as this property has been used in many contexts to prove many results concerning real world network graphs. As an example, we may recollect unweighted sparse graph clustering; an algorithm was designed by Chen et al. [3] specifically for sparse graphs. The limitations of existing sparsity measures become even more apparent for weighted graphs, although most of the real world scenarios result in weighted graphs. Such limitations, in conjunction with the absence of a consensus definition of sparsity, give one an impetus to ponder over the question again about how to compute the sparsity of a network graph. Below, some of the existing measures of sparsity have been recounted.

**Edge Density:** Edge density of a simple undirected graph G= (V, E) is given by $|E|/\binom{|V|}{2}$, which is the ratio of the edges the graph actually has compared to its potential edges [5]. This ratio provides a reflection of how close the graph is to a complete graph, noting that for a complete graph the edge density is 1.

Edge density, being a simple ratio, doesn't take the edge weights into consideration for a weighted graph.

**Degeneracy:** A graph G is $k$-degenerate if for each induced subgraph H of G, $\partial(H) \leq k$, $k$ being a non-negative integer and $\partial(H)$ being the minimum degree of a vertex in H [15]. The concept of k-core of a graph often goes hand in hand with its degeneracy; k-core of a graph being the largest connected subgraph with every vertex having a degree at least k. The degeneracy of a graph is the largest k for which the graph has a non-empty k-core.

Evidently, with sufficiently large k, k-core of a graph represents a highly interconnected community of nodes with each node being connected to at least k other nodes in the community and forms a dense region (which is at the other end of the spectrum than sparsity). Efficient community detection algorithms have been proposed [20] based on reduction of a graph to its k-core subgraph. Extending the concept of k-core to directed graphs and bringing in in-degrees and out-degrees in its definition, community detection algorithms for networks have been devised and quality evaluation of such communities has been proposed [9].

Arboricity and Maximum Average Degree are also used as sparsity measures; however, as they are within a constant factor of degeneracy [18], [6] they are not being treated separately. Moreover, degeneracy can be computed in linear time [6] and hence is a measure popular over many others.

It is important to note that the limiting behavior of a class of graphs is not the concern in this paper and thus we are not focusing our attention on the "sparse class of graphs".

## 3. *Proposed Measure(s)*

As the use of Gini Index is to be explored as a measure of sparsity of network graphs, we would first recapitulate what a Gini Index is.

**Gini Index** – Gini Index [10] measures how equitably a resource is distributed among entities in a population [7]. As it will be seen shortly, this index is normalized and lies between 0 and 1. It measures the extent of inequality in distribution of a certain quantity among certain number of entities. Gini Index is typically computed from the Lorenz curve [16] which plots the cumulative percentage of total quantities, which, a certain fraction of the population can have. In other words, a point on the Lorenz curve denotes the cumulative percentage of total quantity (its vertical-coordinate value) that is allocated to the cumulative fractile of the population (its horizontal-coordinate value). Statements like "20% of the population enjoys 80% of the total allocated fund for the population" can be made using such curves. The graphical definition of Gini Index follows from the Lorenz curve and is equal to twice the area bounded by the Lorenz curve and the line of equality (i.e., the $45^0$ line).

A sparse representation, however, is one in which a small number of entities contain a large proportion of the total quantities. So, a sparse representation inherently contains a large number (compared to the population size) of zero or small-magnitude values such that only a few entities can have significant values. When all the concerned entities receive an equal quantity, the distribution is said to be the least sparse with the Gini Index value equal to zero. The distribution gets sparser as the total quantity gets more and more concentrated only among a few entities. Gini index, essentially a summary statistic, finds wide use in income (or wealth) distribution in economics [14], signal reconstruction from sparse signals in signal and image processing [21], genetics [1].

**Applying Gini Index to compute sparsity of network graphs**: To explore how we can apply Gini Index to determine sparsity of a network graph, let us first consider a simple undirected network graph G= (V, E) representing the interactions among a group of n individuals, i.e, the graph has n nodes with $|V|$=n. Let the corresponding adjacency matrix be given by $A = \left((a_{ij})\right)_{n \times n}$ with $a_{ii}$ being zero $\forall i = 1,2 \ldots, n$ and $a_{ij}$ representing whether an interaction between the nodes $i$ and $j$ has taken place, with $a_{ij} \in \{0,1\}, i,j = 1,2 \ldots, n$.

Let the total number of connections made by the ith individual with others in the graph be given by $a_i$, where $a_i = \sum_{j=1}^{n} a_{ij} \ \forall i = 1,2 \ldots, n$ and the sum of all the connections made by n individuals be denoted by T, where $T = \sum_{i=1}^{n} a_i = \sum_{i=1}^{n} \sum_{j=1}^{n} a_{ij}$.

Let us consider the vector $\underline{a} = [a_1, a_2, \ldots, a_n]$ representing the number of connections made by each of these $n$ individuals. In other words, $\underline{a}$ represents the vector of degrees of the $n$ vertices of the graph.

Let the elements of $\underline{a}$ be arranged in an ascending order and let $b_i$ denote the $i$-th ordered statistic of $\{a_1, a_2, \ldots, a_n\}, i.e., b_1 \leq b_2 \leq \cdots \leq b_n$. It follows that $\sum_{i=1}^{n} b_i = T$.

Using the above notations for a network graph let us now try to draw the Lorenz curve for the vector $\underline{b}$ of ordered node-degrees. The idea here is to see how the sum $T$ of all degrees of $n$ nodes is distributed (relative distribution) among the n nodes of the graph. The fractiles of the order of the graph $n$ will be plotted along the horizontal axis and the cumulative sum of $b_i's$ will be plotted along the vertical axis (Fig 1).

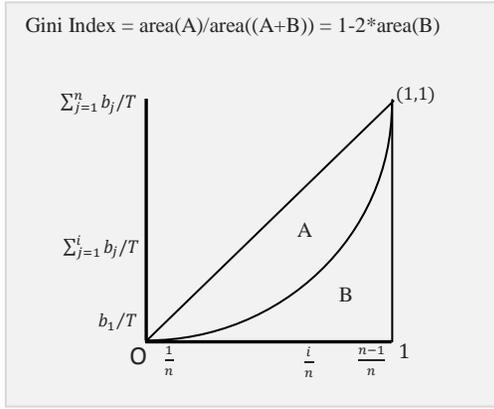

**Fig 1: Gini Index definition using Lorenz Curve for sparsity in distribution of the degrees of nodes of a network graph**

A point on the Lorenz curve, thus drawn, represents the proportion of the total degree of the vertices (y-coordinate) allocated to the specific fractiles (x-coordinate) of the nodes.

To compute Gini Index as indicated in Fig 1 (i.e., Gini Index = area(A/(A + B)) = $1 - 2 * \text{area}(B)$), an estimate of area B needs to be computed. In the literature [8] an estimate of area B has been calculated by approximating the Lorenz curve by piecewise linear segments and the area B approximately equals to the area of:

- a triangle formed by the vertices $(0,0)$, $\left(\frac{1}{n}, 0\right)$ and $\left(\frac{1}{n}, \frac{b_1}{T}\right)$ and
- (n-1) trapeziums – the four vertices of the $i$th. trapezium being:
  $\left(\frac{i}{n}, 0\right), \left(\frac{i+1}{n}, 0\right), \left(\frac{i}{n}, \sum_{j=1}^{i} b_j\right)$ and $\left(\frac{i+1}{n}, \sum_{j=1}^{i+1} b_j\right)$.

Let us represent the area of any polygon by the symbol $\Delta$.

$$\Delta(B) = \Delta\left(triangle\left\{(0,0), \left(\frac{1}{n}, 0\right), \left(\frac{1}{n}, \frac{b_1}{T}\right)\right\}\right) + \Delta((n-1)trapeziums)$$

$$= \frac{1}{2} * \frac{1}{n} * \frac{b_1}{T} + \sum_{i=2}^{n} \frac{1}{2T} * \frac{T}{n} * \left\{\frac{b_1 + b_2 + \cdots + b_{i-1}}{T} + \frac{b_1 + b_2 + \cdots + b_i}{T}\right\}$$

$$= \frac{1}{2nT}[b_1 + \{b_1 + (b_1 + b_2)\} + \{(b_1 + b_2) + (b_1 + b_2 + b_3)\} + \cdots + \{(b_1 + b_2 + \cdots + b_{n-1}) + (b_1 + b_2 + \cdots + b_n)\}]$$

$$= \frac{1}{2nT}[b_n + 3b_{n-1} + 5b_{n-2} + \cdots + \{2(n-2) + 1\}b_2 + \cdots + \{2(n-1) + 1\}b_1]$$

$$= \frac{1}{2nT}[(2n-1)b_1 + (2n-3)b_2 + \cdots + 3b_{n-1} + b_n]$$

$$= \frac{2}{2nT}\left[\left(n - 1 + \frac{1}{2}\right)b_1 + \left(n - 2 + \frac{1}{2}\right)b_2 + \cdots + \left(n - (n-1) + \frac{1}{2}\right)b_{n-1} + \left(n - n + \frac{1}{2}\right)b_n\right]$$

$$= \frac{1}{nT}\sum_{i=1}^{n}\left(n - i + \frac{1}{2}\right)b_i.$$

Therefore, $\text{area}(B) = \frac{1}{2nT}[(2n-1)b_1 + (2n-3)b_2 + \cdots + 3b_{n-1} + b_n]$ and

**Gini Index (GI)** $= 1 - 2.\Delta(B) = 1 - 2.\left[\sum_{i=1}^{n}\frac{b_i}{T}\left(\frac{n - i + \frac{1}{2}}{n}\right)\right]$ ……….. (1)

The formula above is the existing Gini Index formula used in various contexts, signal processing [21] being an example of such a context.

From the above expression of GI, it can be noted that GI is normalized and hence it lies between 0 and 1. A high value of GI (close to 1) indicates that the sum of all vertex-degrees inherent in the graph is centered only among a few individuals whereas when GI is small there are only very few dominant individuals.

**Special Case 1- Complete Graph**: For a complete graph $T = n(n-1)$. The Lorenz curve in this case coincides with the $45^0$ line and the resulting Gini Index is zero. All nodes have equal degrees and hence the sparsity is zero and the result matches our intuition.

**Special Case 2 – Regular cycle**: Each vertex has degree 2 and the above definition of Gini Index yields a value zero. In fact, for a graph in which all the vertices have a constant degree (each less than *n-1*), Gini Index yields zero.

This however does not blend well with the notion of sparsity in graphs. A very large regular cycle can still be sparse, going by the notion of sparsity which is the amount of deviation from the fully connected graph. A regular cycle with 1000 nodes, for example, has 1000 edges with a total degree of 2000; but it has got the potential to have up to 999000 total degrees, implying that the graph is highly sparse. Therefore, it indicates that sparsity of a graph is not just a measure of relative difference in terms of degree of the vertices, but it should be measured with respect to a certain total quantity in terms of degrees. Let that total quantity be $T_1$ with $T_1 \geq T$. Let us now look at how the sparsity measure changes when computed on the basis of $T_1$. In our example of the regular cycle with 1000 nodes, $T=2000$ and $T_1 = 999000$.

We will first prove a lemma to show that the Lorenz curve drawn with respect to $T_1$ meets the vertical line (1,0) to (1,1) at a point lower than (1,1).

**Lemma**: $\frac{\sum_{j=1}^{i} b_j}{T_1} \leq \frac{i}{n}, \forall\, i \leq n.$

**Proof**: $\frac{\sum_{j=1}^{i} b_j}{T_1} = \frac{\sum_{j=1}^{i} b_j}{T} \cdot \frac{T}{T_1} \leq \frac{\sum_{j=1}^{i} b_j}{T}$ $since$ $\frac{T}{T_1} \leq 1$.

Therefore, $\frac{\sum_{j=1}^{i} b_j}{T_1} \leq \frac{i}{n} \Leftrightarrow \frac{\sum_{j=1}^{i} b_j}{T} \leq \frac{i}{n} \Leftrightarrow \frac{b_1+b_2+\cdots+b_i}{b_1+b_2+\cdots+b_n} \leq \frac{i}{n} \Leftrightarrow (n-i)(b_1 + b_2 + \cdots + b_i) \leq i(b_{i+1} + b_{i+2} + \cdots + b_n)$.

But, $(n-i)(b_1 + b_2 + \cdots + b_i) \leq (n-i)ib_i$ $and$ $i(b_{i+1} + b_{i+2} + \cdots + b_n) \geq i(n-i)b_{i+1}$. Hence, it suffices to show that $(n-i).ib_i \leq i.(n-i)b_{i+1}$, which is true anyway since $b_i \leq b_{i+1}$. Thus the claim is proved.

Taking into account the above result, the Lorenz curve is drawn on the basis of $T_1$ as in Fig 2:

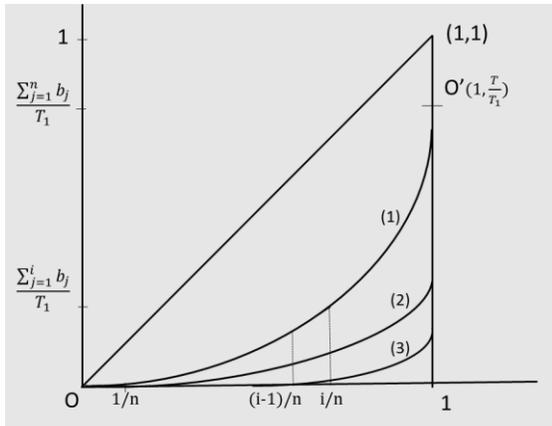

Fig 2: Lorenz curves with different values of $T_1 \geq T$, $T_1$ being a fixed quantity

By the same process of computing the area under the Lorenz curve as has been done previously, using piecewise linear approximation, the revised form of our sparsity measure along the line of Gini Index, say $S(\underline{b})$, is given by

$$S(\underline{b}) = 1 - \frac{1}{nT_1}[(2n-1)b_1 + (2n-3)b_2 + \cdots + 3b_{n-1} + b_n] = 1 - 2.[\sum_{i=1}^{n} \frac{b_i}{T_1}\left(\frac{n-i+\frac{1}{2}}{n}\right)] \quad \ldots\ldots\ldots (2)$$

Let us call $S(\underline{b})$ the Sparsity Index (SI) of a network graph in which the vector $\underline{b}$ stands for the vector of the ordered degrees of its nodes.

With $T_1 = T$, the expressions (1) and (2) are the same. With different choices of $T_1$, the curve $OO'$ takes different forms and the sparsity measure $S(\underline{b})$ changes accordingly.

For computing sparsity of a simple undirected graph, the most plausible choice of $T_1$ is $n(n-1)$, i.e., when it is computed with respect to the total degree of all nodes in the graph. The idea here is to calculate sparsity with respect to the potential total degrees in the graph rather than the actual total degrees.

With the above modification to the expression for Gini Index to construct our sparsity measure $S(\underline{b})$, let us re-visit our two special cases. For **a complete graph**, with $T_1 = n(n-1)$, $S(\underline{b})$ is still zero, which is what it should be. For **a regular cycle** however, if there are n nodes in the cycle, $S(\underline{b}) = (n-3)/(n-1)$. For any regular graph with n nodes and with degree of each node being d (with $d<(n-1)$), $S(\underline{b}) = \frac{n-d-1}{n-1}$.

The two measures as we have defined above, viz., Gini Index and Sparsity Index, of a network graph reveal two different characteristics in front of us, although both of them deal with the degrees of the nodes. Let us examine the relationship between the two measures. Clearly, from expressions 1) and 2) above, it follows that:

$SI = 1 - \frac{1}{nT_1}.S$ and $GI = 1 - \frac{1}{nT}.S$, where $S = 2\sum_{i=1}^{n} b_i (n-i+1/2)$

Therefore, $SI = 1 + \frac{T}{T_1}\left(1 - \frac{1}{nT}.S\right) - \frac{T}{T_1} = 1 - \frac{T}{T_1} + \frac{T}{T_1}GI = \left(1 - \frac{T}{T_1}\right).1 + \frac{T}{T_1}GI$

From the above relationship it follows that – i) SI can be expressed as a convex combination of 1 and GI, noting that $0 < \frac{T}{T_1} \leq 1$ and ii) $SI \geq GI$, the equality being satisfied when $T_1 = T$.

So far we have only dealt with the degrees of the nodes of a graph, i.e., we have concerned ourselves only with the number of connections made by individuals represented by the degrees of the nodes. We must also look at how this expression changes, if at all, for weighted graphs, in which the connection strength or the edge weights are also taken into account.

**Weighted Networks:** As we consider weighted networks, an important distinction must be highlighted – the network graph adjacency matrix now consists of edge weights, i.e., for weighted undirected graphs, $a_{ij}$ now represents the edge weight or the frequency of interaction between the individuals $i$ and $j$ and the constraint $a_{ij} \in \{0,1\}$ no longer exists. What we have instead is $a_{ij} \geq 0$; let us say $a_{ij}$ are non-negative integers, without any loss of generality.

Let us note that for weighted graphs we have two aspects to consider – firstly, the degree of each node representing how well connected that node is with other nodes in the graph and secondly, the edge weights representing the strength or frequencies of such connections. Our treatise so far concerning degrees of nodes remains valid for weighted graphs also for the first case. However, the entries of the adjacency matrix now represent the second case and for this it appears to be intuitively appealing to treat each weighted graph as a multigraph [19].

A multigraph is a graph that can have multiple edges between a pair of nodes. A weighted graph can be treated as a multigraph by taking as many edges between a pair of nodes as the edge-weight between that particular pair of nodes. Treated thus, the adjacency matrix of the multigraph remains the same as the adjacency matrix of the weighted graph; but the edge weights get added to the degrees of the corresponding vertices. Hence we still have $a_i = \sum_{j=1}^{n} a_{ij} \; \forall i = 1,2 \dots, n$, even though $a_{ij}$'s represent the edge weights in the weighted network. Once the mapping between the edge weights of a network graph to the vertex-degrees of the corresponding multigraph is established, the rest of the treatment remains the same as in the case of a simple unweighted network. However it may be noted that the treatment mentioned above is valid since $a_{ij}$'s are considered to be non-negative integers.

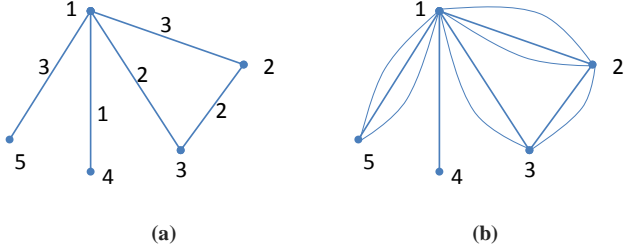

(a)          (b)

**Fig 3: (a) weighted network (b) multigraph equivalent to weighted graph, both having the same adjacency matrix**

For weighted graphs, $a_{ij}$ can be greater than 1, and a few other choices of $T_1$ can be considered; examples being $T_1 = n(n-1).max\,(a_{ij})$ and $T_1 = n.b_n$. Clearly, the first one is one of the extreme cases and sparsity will be very high if it is measured with respect to such an extreme value. The second choice of $T_1$, however, is perhaps more realistic in which sparsity is measured against the total edge weight obtained by using the maximum edge weight associated with any node in the network. In this case also the sparsity is expected to be high, but not as much as in the previous case. The corresponding Lorenz curves for these two cases have been indicated by (3) and (2) respectively in Fig. 2 (with $T_1 = n(n-1).max(a_{ij})$ and $T_1 = n.b_n$ respectively). The choice of $T_1$ is guided by the problem under consideration and the basis on which one wishes to measure sparsity of the network graph.

The concept of measuring sparsity with Gini Index can be extended for directed graphs also.

### *4. Properties of Sparsity Measure*

Now that our sparsity measure has been defined, we would like to see how it behaves with respect to the properties of a sparsity measure. For our purpose we will follow the criteria used by Hurley and Rickard [13] in their comparison of fifteen different sparsity measures showing their relative merits and demerits. Six different criteria have been used by them for comparing different measures of sparsity, including Gini index, and according to their analysis Gini Index comes out to be the only one to satisfy all the six criteria (viz., Robin Hood, Scaling, Rising Tide, Cloning, Bill Gates and Babies). It will be checked how these properties work out for the sparsity index as a sparsity measure. For this purpose, let us assume $T_1 = n(n-1)$ as has been done earlier for a simple undirected graph, unless stated otherwise.

1. D1 Robin Hood: Taking a quantity from a rich and giving it to a poor decreases sparsity, i.e., $S([b_1 \dots b_i + \alpha \dots b_j - \alpha \dots b_n]) < S(\underline{b})$ for all $\alpha, b_i, b_j$ such that $0 < \alpha < \frac{b_j - b_i}{2}$, where S denotes a sparsity measure.
   This property is satisfied for our definition of $S(\underline{b})$.
   Let $\underline{\acute{b}} = [b_1 \dots, b_i + \alpha, \dots, b_j - \alpha, \dots b_n]$ and $\underline{b} = [b_1 \dots, b_i, \dots, b_j, \dots b_n]$.
   Then, $S(\underline{\acute{b}}) - S(\underline{b}) = -\frac{2}{nT_1} \left[ \alpha \left( n - i + \frac{1}{2} \right) - \alpha \left( n - j + \frac{1}{2} \right) \right] = -\frac{2\alpha}{nT_1}(j - i)$. But $j > i$ implies $\frac{2\alpha}{nT_1}(j - i) > 0$.
   Therefore, $S(\underline{\acute{b}}) - S(\underline{b}) < 0$ and the property holds.

2. D2 Scaling: Multiplication by a constant factor does not change sparsity, i.e., $S(\alpha \underline{b}) = S(\underline{b}) \; \forall \; \alpha > 0, \alpha \; real, \alpha \neq 1$.
   In general, computing $S(\alpha \underline{b}) - S(\underline{b})$ we get $\frac{2(1-\alpha)}{nT_1} \sum_{i=1}^{n} b_i \left( n - i + \frac{1}{2} \right)$ after simplification.

Therefore, $S(\alpha\underline{b}) < S(\underline{b})$ for for $\alpha > 1$, i.e., sparsity decreases for $S(\alpha\underline{b})$ compared to $S(\underline{b})$ with $\alpha > 1$ and increases for $\alpha < 1$.

However it can be verified that scaling is invariant if each $a_{ij}$ is multiplied by $\alpha$ for our two cases $i)\ T_1 = n(n-1).max a_{ij}$ and $ii)\ T_1 = n.b_n$ as $\alpha$ gets cancelled out in both the cases in the $b_i/T_1$ term in the expression of $S(\alpha\underline{b})$.

Let's consider the first case. As each $a_{ij}$ is multiplied by $\alpha \neq 1$, so does $max a_{ij}$ and also, $\sum_{j=1}^{n} \alpha.a_{ij} = \alpha.\sum_{j=1}^{n} a_{ij} = \alpha.a_i$. Therefore, $S(\alpha\underline{b}) = \frac{2}{n.n(n-1)\alpha.\max(a_{ij})} \sum_{i=1}^{n} \alpha b_i \left(n - i + \frac{1}{2}\right) = S(\underline{b})$. The other case works out in a similar manner.

3. **D3 Rising Tide:** $S(\alpha + \underline{b}) < S(\underline{b})$, $\forall\ \alpha > 0, \alpha\ real$ and all $b_i$'s are not equal.

   Computing $S(\alpha + \underline{b}) - S(\underline{b})$ yields $\left(-\frac{2\alpha}{nT_1}\right).\sum_{i=1}^{n}\left(n - i + \frac{1}{2}\right) = \left(-\frac{\alpha}{T_1}\right).n$ after simplification. This expression is evidently negative as all quantities involved in this expression are positive quantities. Hence the inequality holds for our case.

4. **D4 Cloning:** $S(\underline{b}) = S(\underline{b}\|\underline{b}) = S(\underline{b}\|\underline{b}\|\underline{b}) = \cdots$ where '$\|$' denotes the concatenation operator for two vectors.

   The following counter-example shows that our sparsity measure is not invariant under cloning. Let us consider a 4-node unweighted cycle represented by the degree vector (2,2,2,2). The sparsity measure when computed on the basis of expression (2) yields 0.33, taking $T_1 = n(n-1) = 12$. As the concatenation operator produces an 8-node cycle the value of the sparsity measure with $T_1 = n(n-1) = 56$ now yields 0.71. The sparsity of the higher degree cycle is expectedly more, as the degree-vector is self-concatenated in this case. Similar logic holds for the case of weighted graphs.

5. **P1 Bill Gates:** $\forall i\ \exists\ \beta_i = \beta$ such that $S(b_1, \ldots, b_i + \beta + \alpha \ldots b_n) > S(b_1, \ldots, b_i + \beta \ldots, b_n)\ \forall\ \alpha > 0$.
   This property is satisfied for our definition of $S(\underline{b})$.

   After simplifying $S(b_1, \ldots, b_i + \beta + \alpha \ldots b_n) - S(b_1, \ldots, b_i + \beta \ldots, b_n)$, we get $\alpha.\left(n - i + \frac{1}{2}\right)$. Since each of the quantities involved in this expression are non-negative and $1 \leq i \leq n$, $\alpha.\left(n - i + \frac{1}{2}\right) > 0$ and the inequality follows.

6. **P2 Babies:** $S(\underline{b}\|0) > S(\underline{b})$.
   Adding a group of isolated nodes (zero degrees/weights) to a network increases its sparsity and that is true for our case also.

Therefore, our sparsity measure directly satisfies four of the six criteria, does not satisfy cloning and satisfies scaling for specific values of $T_1$.

**Comparison with Edge Density:** We will first prove a theorem relating sparsity index with edge density.

**Theorem 1:** As the edge density of a graph increases its sparsity index decreases, i.e., edge density and sparsity index follow the same trend for a changing graph.

**Proof:** Suppose a simple graph G has n nodes and k edges. Let us first show that as k is increased to (k+1), SI decreases, i.e, if H be the graph with n nodes obtained after adding an edge to G then SI(G) > SI(H). Let us take $T_1 = n(n-1)$ in the expression of SI, as the graphs under consideration are simple graphs.

Let $\underline{b} = (b_1, b_2, \ldots, b_n)$ be the degree sequence of the graph G and $\underline{c} = (c_1, c_2, \ldots, c_n)$ be the degree sequence of the graph H. It may be noted that $\sum_{i=1}^{n} b_i = 2k$ and $\sum_{i=1}^{n} c_i = 2(k+1)$.

Now, $SI(G) \geq SI(H) \Leftrightarrow 1 - \frac{2}{n^2(n-1)}\sum_{i=1}^{n} b_i(n-i+1/2) \geq 1 - \frac{2}{n^2(n-1)}\sum_{i=1}^{n} c_i(n-i+1/2)$

$\Leftrightarrow \sum_{i=1}^{n}(2n-2i+1)(c_i - b_i) \geq 0 \Leftrightarrow (2n+1)\sum_{i=1}^{n}(c_i - b_i) + 2\sum_{i=1}^{n}(ib_i - ic_i) \geq 0$

$\Leftrightarrow (2n+1).2 + 2\sum_{i=1}^{n}(ib_i - ic_i) \geq 0$, since $\sum_{i=1}^{n}(c_i - b_i) = 2$

$\Leftrightarrow 2n+1 + \sum_{i=1}^{n}(ib_i - ic_i) \geq 0$ --------------------------(i)

We note that the graph G under consideration cannot be complete, as, in that case the question of adding another edge to G does not arise. Hence we assume that G is short of being a complete graph, i.e., $k < \frac{n(n-1)}{2}$.

$\underline{b}$ and $\underline{c}$ being ordered degree sequences, $b_1 \leq b_2 \leq \cdots \leq b_n$ and $c_1 \leq c_2 \leq \cdots \leq c_n$. Let us also note that the degree sequence $\underline{c}$ is obtained from $\underline{b}$ as a consequence of adding one edge to G, i.e., except for that one edge the two graphs G and H are the same. Therefore, the degrees of only the two nodes joining that additional edge will be different in the two graphs; the rest of the nodes will have the same degrees.

Let $c_p = b_p + 1, c_q = b_q + 1$ and $c_i = b_i \ \forall \ i = 1,2,\ldots,n \ and \ i \neq p,q$. Let $p < q$.

$\sum_{i=1}^{n}(ib_i - ic_i) = (b_1 + 2b_2 + \cdots + pb_p + \cdots + qb_q + \cdots + b_n) - (b_1 + 2b_2 + \cdots + p(b_p + 1) + \cdots + q(b_q + 1) + \cdots + b_n) = -(p+q)$. Substituting this value in the expression (i), we get

$SI(G) \geq SI(H) \ iff \ 2n+1 - (p+q) \geq 0$. But, (p+q) can at most take a value of (n+n-1)=2n-1.

Therefore, the above inequality satisfies and we are able to show that if the edge density is increased (by $2/n(n-1)$ in this particular case) in a graph, its sparsity decreases. The result can be generalized in a similar manner if more than one edge is added to the graph.

Remarks: 1) Adding the degrees to the nodes towards the higher end of the ordered sequence, such that the sequence can still generate a valid graph, produces SI values towards the maximum.

2) Given n and k, a graph with degree distribution having maximum sparsity can also be obtained in a similar manner. ∎

Evidently, edge density is a simple ratio and is nothing but $T/T_1$ for the simple graph with $T_1 = n(n-1)$. This is represented by a point on the Lorenz curve ($O'$, with height of $O'$ being $T/T_1$) as shown in Fig 2. Therefore, Sparsity Index, despite being a summary index, is a much more fine-grained measure compared to a simple ratio.

Below is an example of two simple graphs with the same edge density, but the sparsity measure $S$ yields different values in the two cases.

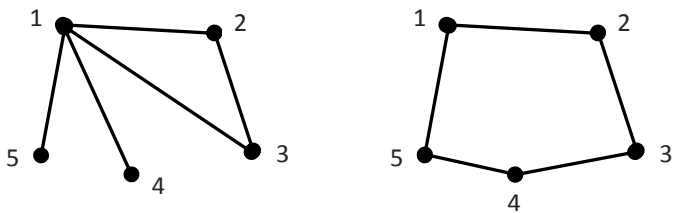

Fig 4: Graphs with same edge density (value = 0.5), but with different values of Sparsity Index (respectively 0.64 and 0.5 taking $T_1 = n(n-1) = 20$) and Gini Index (respectively 0.28 and 0).

A higher value of Sparsity measure indicates more variation among the node-degrees; for the first figure we have $\underline{b} = [4,2,2,1,1]$ and for the second $\underline{b} = [2,2,2,2,2]$; evidently the first graph has a higher variation of its node-degrees and that explains the result as indicated in Fig 4.

One can also construct another set of two graphs with $\underline{b}$ = [4,3,2,2,1] and [3,3,2,2,2], the corresponding sparsity indices being 0.54 and 0.46 respectively (and Gini indices being 0.23 and 0.1 respectively). These two graphs are obtained by adding an edge each to the previous two graphs, thus the edge densities remain the same, but the first graph shows a higher variation among the node-degrees reflected in its higher sparsity index compared to the second. Also, compared to the previous set of graphs, these two graphs have higher edge densities showing a decrease in value of the sparsity indices.

It follows from the above discussion that sparsity of a network graph represents a distribution of the total strength of all interactions in the graph among the constituent nodes. If only a few nodes show strong interactions in the graph, we are expected to have a high sparsity index; on the contrary, if there are many nodes with moderate interaction strengths we are expected to achieve a relatively low value of the sparsity index.

Also, going by a popular interpretation of the node-degrees, especially in social networks, which is the amount of influence an individual carries in the network, a high value of Gini Index may indicate the presence of a few strongly influential individuals in the network.

Notably, edge density falls especially short in the case of weighted graphs, as this measure does not take into account the weights. But real life networks are represented in most cases by weighted graphs to indicate tie-strength. Gini Index, when computed on the basis of the edge weights of a network graph, can indicate how sparse the distribution of total weights is among the constituent nodes of the network and thus, provide us with a first impression of the nature of connectedness of the individuals in the network, whether a small group of individuals are intimately connected or a large number of them rather have occasional interactions. Also, if clusters exist in the network, the corresponding GI value is expected to be high. On the other hand, if the GI value is small it is unlikely that the network produces good clusters.

**Computational Complexity** - Both GI and SI can be calculated easily even for large graphs, as, computationally this can be done in $O(nlog(n))$ time, the ranking of the nodes in terms of degrees being the main contributing factor towards complexity.

We are interested in examining now if there is any relationship between the sparsity index and the distribution of the degrees of nodes of a graph.

## 5. *Degree distribution and sparsity of a network graph*

While commenting on the characterization of complex interaction networks, Barabasi et al. [2] point out that the probability that a vertex in the network interacts with $x$ other vertices in the network decays as a power law; attributing the reason primarily to two factors, viz., growth and preferential attachment. Later, Clauset et al. [4] present a statistical analysis of goodness of fit, using methods postulated by them, of as many as twenty-four real world large network datasets to power law. Their analysis reveals that for nearly a third of these datasets, there is less than 5% probability of a power law distribution being followed, as evidenced from the data, defeating the conjecture of a power law distribution in all these cases. On the contrary, a good number of the datasets provides evidence to support a power law conjecture. Muchnik et al. [17] reinforce that the probability distribution of number of ties of an individual in a social network follows a scale-free power law and go on to explain the network formation in relationship with the volume of a user's activities in the network.

In our attempt to link sparsity with the degree distribution of a network graph, we therefore take the degree distribution to follow power law which is given by $p(x) = C.x^{-\beta}$ with the exponent $\beta > 1$. In the context of a graph $p(x)$ is being defined in the following way.

For a graph $G = (V, E)$ with $|V| = n$ and $|E| = N$, let the probability that a node in G has degree $x$ be denoted by $p(x)$ where x ≥ 1 is a non-negative integer. Let the maximal degree of the graph $G$ be denoted by $k$. In our setting, $b_1$ represents the node with minimum number of connections, $b_2$ the second minimum and so on and finally $b_n = k$.

To explore the nature of the relationship between a degree distribution that follows power-law and the sparsity of a graph, we will construct a synthetic example as below. From this example, we will first check how the power-law degree distribution relates with the sparsity as expressed by Gini index and subsequently by the sparsity index.

The below table gives us the way we are to compute the degree frequency distribution of network graphs:

Table 3: Constructing Degree Frequency Distributions

| Degrees (i) | $p_{1i}$ | $p_{2i}$ |
|---|---|---|
| 1 | $C_1$ | $C_2$ |
| 2 | $C_1 \cdot 2^{-\beta_1}$ | $C_2 \cdot 2^{-\beta_2}$ |
| . | . | . |
| . | . | . |
| . | . | . |
| k | $C_1 \cdot k^{-\beta_1}$ | $C_2 \cdot k^{-\beta_2}$ |
| Total | $C_1 \sum_{i=1}^{k} i^{-\beta_1} = 1$ | $C_2 \sum_{i=1}^{k} i^{-\beta_2} = 1$ |

We will make use of the above table in constructing our example distributions and the discussion that follows.

**Example**: For a wide variety of interaction networks characterized by power law degree distribution, it has been shown empirically that the exponent term $\beta$ typically lies between 2 and 3 [4]. Following this observation, in our construction of an example distribution we will take 3 different values of $\beta$ in the range between 2 and 3 and one value outside (below 2). We represent the various degree values by $i$ and the frequencies, i.e., the number of nodes having that degree value, by $f_i, i = 1, 2, \ldots, k$.

Table 1: The below table contains degree frequency distributions of three graphs following power law degree distribution with exponents 2, 2.5 and 2.75 respectively. The graphs are of equal sizes and have the same highest degree (n=200 and k=11 for all three).

| Degree Frequency Distr with $\beta = 2$ ||| Frequency Distr with $\beta = 2.5$ ||| Frequency Distr with $\beta = 2.75$ |||
|---|---|---|---|---|---|---|---|---|
| $i$ | $p_i = C_1 \cdot i^{-\beta_1}$ | $f_i = n p_i$ | $i$ | $q_i = C_2 \cdot i^{-\beta_2}$ | $f_i$ | $i$ | $r_i$ | $f_i$ |
| 1 | 0.64 | 128 | 1 | 0.75 | 150 | 1 | 0.8 | 160 |
| 2 | 0.16 | 32 | 2 | 0.13 | 26 | 2 | 0.12 | 23 |
| 3 | 0.07 | 14 | 3 | 0.05 | 10 | 3 | 0.04 | 8 |
| 4 | 0.04 | 8 | 4 | 0.02 | 4 | 4 | 0.009 | 2 |
| 5 | 0.02 | 4 | 5 | 0.02 | 3 | 5 | 0.008 | 1 |
| 6 | 0.02 | 4 | 6 | 0.008 | 2 | 6 | 0.008 | 1 |
| 7 | 0.01 | 2 | 7 | 0.008 | 1 | 7 | 0.005 | 1 |
| 8 | 0.01 | 2 | 8 | 0.007 | 1 | 8 | 0.003 | 1 |
| 9 | 0.01 | 2 | 9 | 0.003 | 1 | 9 | 0.003 | 1 |
| 10 | 0.01 | 2 | 10 | 0.002 | 1 | 10 | 0.002 | 1 |
| 11 | 0.01 | 2 | 11 | 0.002 | 1 | 11 | 0.002 | 1 |
| Total | $\sum_{i=1}^{k} p_i = 1$ | 200 | | 1 | 200 | | 1 | 200 |

Edge densities of the graphs corresponding to the above frequency distributions are 0.01, 0.008 and 0.007 respectively.

The constant $C$ for each of the three distributions is computed from the equation: $C . \sum_{i=1}^{k} i^{-\beta} = 1$ and the rest of the figures are calculated based on the formulae as indicated in the first frequency table (with $\beta = 2$).

We will construct another degree frequency distribution with $\beta = 1.7$, for which the $f_i$ values can be calculated as: {112, 34, 18, 10, 6, 6, 4, 4, 2, 2, 2}. The edge density for the corresponding graph is 0.011.

It is important that we ensure that the above degree sequences are indeed capable of producing valid graphs. To test the validity we have applied Havel-Hakimi algorithm [11] on each of the degree sequences, generated from the degree frequency distribution, and have come to a positive conclusion in each case.

We will now draw the frequency distribution curves and also the Lorenz curves corresponding to the distributions with four different $\beta$-values (Fig 5).

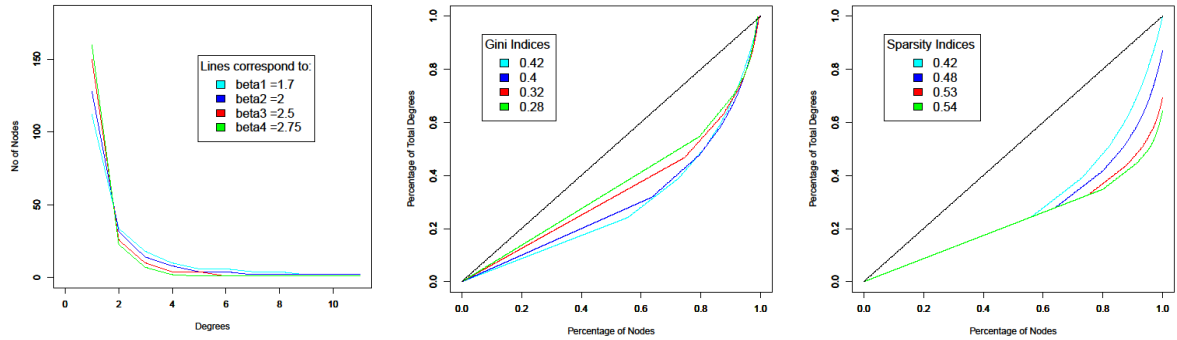

**Figure 5: Frequency Plot and Lorenz curves for Power Law degree distribution example as represented by Table 1.**

The first plot represents the usual frequency distribution, the second one shows the Lorenz curves corresponding to each of the above distributions computed on the basis of Gini Index (formula (1) earlier in the document) and the third plot represents the Lorenz curve based on Sparsity Index $S(\underline{b})$ (formula (2), with $T_1 = \max \left( \sum_{i_j=1}^{k} i_j f_{i_j} , j = 1,2,3,4 \right) = 460$ in our example achieved for $\beta = 1.7$ ). The considered value of $T_1$ is the minimum possible value that $T_1$ can take as it is imperative for us to take the same value of $T_1$ for all the four distributions. It is also to be noted that $T_1$ can take other values as indicated in section 3. We represent the various values of the sparsity indices for different choices of $T_1$ in the table below:

Table 2: Sparsity Indices for different choices of $T_1$:

| Exponents | Sparsity Indices with $T_1 = n(n-1)$ | Sparsity indices with $T_1 = nb_n$ |
|---|---|---|
| $\beta = 1.7$ | 0.9932 | 0.88 |
| $\beta = 2$ | 0.9939 | 0.89 |
| $\beta = 2.5$ | 0.9945 | 0.90 |
| $\beta = 2.75$ | 0.9947 | 0.904 |

As shown in the second column of the above table, sparsity indices for all exponent values are close to 1 for the maximum possible choice of $T_1$. The Lorenz curves in this case almost coincide with the horizontal axis. For the other choice of $T_1$ however, the sparsity indices are high, in the neighborhood of 0.9, still lower than the previous case. The corresponding Lorenz curves are as follows:

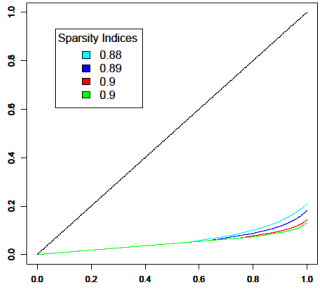

**Figure 6: Lorenz curves showing high sparsity values with $T_1 = nb_n$ in our example**

The two plots of Lorenz curves as in Fig 5 reveal an interesting trend: the sparsity index shows a decreasing trend with respect to the exponent when relative difference among values in the distribution is the concern, whereas, when computed to compare values in the distribution with respect to a fixed constant, larger than the total values involved, the trend is reversed and the sparsity increases with increasing exponent of the underlying power law distribution.

A close scrutiny of Table 1 reveals that of the four graphs corresponding to the four different exponents, the one with the lowest value of the exponent $\beta$ has got the highest number of edges (which is nothing but half of the total degree of all nodes, $\sum_{i=1}^{k} i f_i$, in the graph) thereby indicating the lowest sparsity which matches with our result. On the other hand, the graph with the lowest exponent has got more number of higher degree nodes than others indicative of higher variability; a possible reason of having the highest value of the Gini Index.

This leads us to prove the following theorem describing the relationship between our chosen measure of sparsity and the power law degree distribution.

**Theorem 2:** As the exponent $\beta$ increases the sparsity (based on the Lorenz curve and sparsity index) increases.

**Proof:** Let $p_1(x) = C_1 \cdot x^{-\beta_1}$ and $p_2(x) = C_2 \cdot x^{-\beta_2}$ with $1 < \beta_1 < \beta_2$ denote the power law degree distributions corresponding to two different network graphs. We have to show that the sparsity of the network as represented by $p_2$ is higher than the sparsity of the network as represented by $p_1$ when the sparsity is measured with respect to the same constant $T_1$ larger than the individual total degrees of the graphs.

As $\beta_1 < \beta_2$, $\sum_{i=1}^{k} i^{-\beta_1} > \sum_{i=1}^{k} i^{-\beta_2}$. But $C_1 \sum_{i=1}^{k} i^{-\beta_1} = 1 = C_2 \sum_{i=1}^{k} i^{-\beta_2}$. Therefore, $C_1 < C_2$ ………..(3)

If we can show that $\forall j \geq 1$, $C_1 \cdot \sum_{i=1}^{j} i^{-\beta_1} \leq C_2 \cdot \sum_{i=1}^{j} i^{-\beta_2}$ ………………….. (4), then our conjecture can be proved.

For $j = 1$, the statement (4) is true anyway as $C_1 < C_2$ according to statement (3).

For $j = k$ however, equality holds in statement (4).

For $1 < j < k$, let us say j=r, we have:

$C_1 \cdot \sum_{i=1}^{j} i^{-\beta_1} < C_2 \cdot \sum_{i=1}^{j} i^{-\beta_2} \Leftrightarrow C_1(1 + 2^{-\beta_1} + \cdots + r^{-\beta_1}) < C_2(1 + 2^{-\beta_2} + \cdots + r^{-\beta_2})$

$\Leftrightarrow \frac{1 + 2^{-\beta_1} + \cdots + r^{-\beta_1}}{\sum_{i=1}^{k} i^{-\beta_1}} < \frac{1 + 2^{-\beta_2} + \cdots + r^{-\beta_2}}{\sum_{i=1}^{k} i^{-\beta_2}}$, substituting values of $C_1$ and $C_2$.

$\Leftrightarrow (1 + 2^{-\beta_1} + \cdots + r^{-\beta_1})(1 + 2^{-\beta_2} + \cdots + k^{-\beta_2}) < (1 + 2^{-\beta_2} + \cdots + r^{-\beta_2})(1 + 2^{-\beta_1} + \cdots + k^{-\beta_1})$

$\Leftrightarrow (1 + 2^{-\beta_1} + \cdots + r^{-\beta_1})((r+1)^{-\beta_2} + \cdots + k^{-\beta_2}) < (1 + 2^{-\beta_2} + \cdots + r^{-\beta_2})((r+1)^{-\beta_1} + \cdots + k^{-\beta_1})$ ……(5)

Let us note that $(r+1)^{-\beta_2} + \cdots + k^{-\beta_2} < (r+1)^{-\beta_1} + \cdots + k^{-\beta_1}$ as $\beta_2 > \beta_1$.

Also, $2^{-\beta_1} \cdot (r+1)^{-\beta_2} < 2^{-\beta_2} \cdot (r+1)^{-\beta_1}$ since $2^{\beta_2 - \beta_1} < (r+1)^{\beta_2 - \beta_1}$ and by a similar logic this is true for other terms, i.e., $2^{-\beta_1} \cdot i^{-\beta_2} < 2^{-\beta_2} \cdot i^{-\beta_1}, \forall i = (r+2), \ldots, k$.

Therefore, the inequality in (5) is satisfied and the statement (4) is true for $j = r, 1 < r < k$. Thus the theorem can be proved by the method of induction, the outline of which has been given above. ∎

## 6. Conclusion and Future Works

In this article our main objective is to quantify sparsity so that we can use it in algorithms which deal specifically with sparse network graphs. In this regard, we have defined a new measure of sparsity which is a modification of the Gini Index. Edge Density as a measure of sparsity falls short on many counts; the main one being the content of information it takes into account in its calculation. In arriving at our measure of sparsity, we wanted to address this aspect.

In this article a sparsity index is defined for network graphs, which is in fact a generalization of Gini Index. We have shown that the proposed sparsity index satisfies many of the general properties of a sparsity measure. It is also seen that the proposed sparsity index decreases when edge density increases. Examples are given where edge density values for two different graphs are the same but the sparsity indices are different, thereby indicating that sparsity index provides more information about the graph than edge density. Additionally, we have derived a relationship between the exponent of the power law distribution (which is used to approximate the degree distribution in network graphs) and sparsity index. As the exponent increases the sparsity index increases when computed on the basis of a large constant (larger than the total degrees of the participating graphs of equal sizes and the same maximum degree). This result provides the connection between power law distribution and sparsity index.

In order to make any sparsity measure (be it edge density or the proposed sparsity index) effective, relationships between the sparsity measure and the other properties of the network graph are to be established mathematically. It may be noted that several parameters are calculated from network graphs, like centrality measures, clustering coefficients etc. There is also a large amount of literature where the degree distribution is approximated by power law and other such distributions. Nevertheless, the relationships between these distributions and the parameters have not been adequately explored. Probably the proposed sparsity index may help in bridging this gap. For example, if the network graph is complete, then every node is a central node and the sparsity index value is zero. Though this is an extreme example, relationship between increase in the value of the sparsity measure and several parameters of a network graph with respect to the corresponding degree distribution need to be explored analytically.

Another possible use is to define a bound on the sparsity index to express mathematically when a graph is sparse. The bound need not be a hard bound. For example, if the sparsity measure value is less than a value, say $a$, we may say that the graph is not sparse; if it is greater than a number, say $b, a < b$, we say that the graph is sparse; and if it lies in between $a$ and $b$ then it may need further analysis. Sequential probability ratio test (SPRT) in statistics is one such example where decisions are taken in a step by step manner. The values of $a$ and $b$ may also be determined statistically (not deterministic). This is a possible direction in which further work can be done.

# References

[1] Araya, C.L., Cenik, C., Reuter, J.A., Kiss, G., Pande, V.S., Snyder, M.P. and Greenleaf, W.J., 2016. Identification of significantly mutated regions across cancer types highlights a rich landscape of functional molecular alterations. *Nature genetics*, *48*(2), pp.117-125.

[2] Barabási, A.L. and Albert, R., 1999. Emergence of scaling in random networks. *science*, *286*(5439), pp.509-512.

[3] Chen, Y., Sanghavi, S. and Xu, H., 2012. Clustering sparse graphs. In *Advances in neural information processing systems* (pp. 2204-2212).

[4] Clauset, A., Shalizi, C.R. and Newman, M.E., 2009. Power-law distributions in empirical data. *SIAM review*, *51*(4), pp.661-703.

[5] Diestel, R., 2000. *Graph theory {graduate texts in mathematics; 173}*. Springer-Verlag Berlin and Heidelberg GmbH & amp.

[6] Eppstein, D., Löffler, M. and Strash, D., 2010, December. Listing all maximal cliques in sparse graphs in near-optimal time. In *International Symposium on Algorithms and Computation* (pp. 403-414). Springer Berlin Heidelberg. *arXiv:1006.5440v1* [cs.DS] 28-Jun-2010

[7] Farris, F.A., 2010. The Gini index and measures of inequality. *American Mathematical Monthly*, *117*(10), pp.851-864.

[8] Gastwirth, J.L., 1972. The estimation of the Lorenz curve and Gini index. *The Review of Economics and Statistics*, *54*(3), pp.306-316.

[9] Giatsidis, C., Thilikos, D.M. and Vazirgiannis, M., 2011, December. D-cores: Measuring collaboration of directed graphs based on degeneracy. In *2011 IEEE 11th International Conference on Data Mining* (pp. 201-210). IEEE. doi:10.1109/ICDM.2011.46

[10] Gini, C., 1912. Italian: Variabilità e mutabilità. *Variability and Mutability', C. Cuppini, Bologna*.

[11] Hakami, S., 1962. On the realizability of a set of integers as degrees of the vertices of a linear graph. *SIAM Journal Applied Mathematics*. *10*(3) pp.496–506.

[12] Huang, Z., Chen, H. and Zeng, D., 2004. Applying associative retrieval techniques to alleviate the sparsity problem in collaborative filtering. *ACM Transactions on Information Systems (TOIS)*, *22*(1), pp.116-142.

[13] Hurley, N. and Rickard, S., 2009. Comparing measures of sparsity. *IEEE Transactions on Information Theory*, *55*(10), pp.4723-4741.

[14] Liberato, P. and Lorenzo, G., 2006. Inequality Analysis: The Gini Index. Food and Agriculture Organization of the United Nations.

[15] Lick, D.R. and White, A.T., 1970. k-Degenerate graphs. *Canadian J. of Mathematics*, *22*, pp.1082-1096.

[16] Lorenz, M.O., 1905. Methods of measuring the concentration of wealth. *Publications of the American statistical association*, *9*(70), pp.209-219.

[17] Muchnik, L., Pei, S., Parra, L.C., Reis, S.D., Andrade Jr, J.S., Havlin, S. and Makse, H.A., 2013. Origins of power-law degree distribution in the heterogeneity of human activity in social networks. *Scientific Reports*, 3, article No 1783 doi:10.1038 *arXiv preprint arXiv:1304.4523*.

[18] Nešetřil, J. and de Mendez, P.O., 2012. *Sparsity: graphs, structures, and algorithms* (Vol. 28). Springer Science & Business Media, ISSN 0937-5511

[19] Newman, M.E., 2004. Analysis of weighted networks. *Physical review E*, *70*(5), p.056131.

[20] Peng, C., Kolda, T.G. and Pinar, A., 2014. Accelerating community detection by using k-core subgraphs. *arXiv preprint arXiv:1403.2226*.

[21] Rickard, S. and Fallon, M., 2004, March. The Gini index of speech. In *Proceedings of the 38th Conference on Information Science and Systems (CISS'04)*.

[22] Zonoobi, D., Kassim, A.A. and Venkatesh, Y.V., 2011. Gini index as sparsity measure for signal reconstruction from compressive samples. *IEEE Journal of Selected Topics in Signal Processing*, *5*(5), pp.927-932.